# Aharonov-Bohm oscillations in Dirac semimetal $Cd_3As_2$ nanowires

Li-Xian Wang[1,†], Cai-Zhen Li[1,†], Da-Peng Yu[1,2,3] & Zhi-Min Liao[1,2*]

[1]State Key Laboratory for Mesoscopic Physics, Department of Physics, Peking University, Beijing 100871, China

[2]Collaborative Innovation Center of Quantum Matter, Beijing, China

[3]Institute of Physics and Electronic Information, Yunnan Normal University, Kunming 650500, China

†These authors contributed equally to this work.

*email: liaozm@pku.edu.cn

**Three-dimensional Dirac semimetals, a three-dimensional analogue of graphene, are unusual quantum materials with massless Dirac fermions, which can be further converted to Weyl fermions by breaking time reversal or inversion symmetry. Topological surface states with Fermi arcs are predicted on the surface and have been observed by angle-resolved photoemission spectroscopy experiments. Although the exotic transport properties of the bulk Dirac cones have been demonstrated, it is still a challenge to reveal the surface states via transport measurements due to the highly conductive bulk states. Here, we show Aharonov-Bohm oscillations in individual single-crystal $Cd_3As_2$ nanowires with low carrier concentration and large surface-to-volume ratio, providing transport evidence of the surface state in three-dimensional Dirac semimetals. Moreover, the quantum transport can be modulated by tuning the Fermi level using a gate voltage, enabling a deeper understanding of rich physics residing in Dirac semimetals.**

The Dirac fermions in graphene and topological insulators are confined in a 2D plane, while in a 3D Dirac semimetal the Dirac fermions are extended to the 3D space[1-17]. Recently, $Cd_3As_2$ and $Na_3Bi$ have been predicted to be Dirac semimetals[5,6], which were further identified by angle-resolved photoemission spectroscopy (ARPES) experiments[7-10]. The Dirac point in 3D Dirac semimetals is constituted by two degenerated Weyl nodes and this degeneracy can be lifted by breaking time reversal or inversion symmetry[5,6]. Owing to its unique bulk Dirac cones, the 3D Dirac semimetal has become an excellent platform to investigate exotic transport properties, such as giant positive magnetoresistance (MR)[11], quantum linear MR[5,12], splitting of Landau levels[13-16], and chiral anomaly induced negative MR[17-21]. On the other hand, nontrivial surface states with arc-like Fermi surface are theoretically predicted and observed by ARPES experiments[5,6,22,23]. Nevertheless, the nontrivial surface states are not accessible by transport measurements in bulk Dirac semimetals because of the bulk-dominant conduction.

The Aharonov-Bohm (A-B) effect of individual nanowires with large surface to volume ratio provides an effective method to explore the transport properties of the nontrivial surface states. As the mean free path of the carriers is comparable with the perimeter of the nanowire, the electron wave is confined in the finite boundary condition along the nanowire perimeter. Consequently, the surface energy bands are enforced into discrete subbands due to the quantum confinement[24,25]. As sweeping the magnetic field parallel to the nanowire direction, the density of states at the Fermi level is periodically altered as each subband crosses the Fermi energy[24,25], and the oscillating period is described by the $\Phi/\Phi_0$, where $\Phi$ is the magnetic flux enclosed by the path of the surface carriers and $\Phi_0 = h/e$. These quantum oscillations have been observed in topological insulator nanoribbons and nanowires, which are still called as A-B oscillations in the recent literatures[26-28]. Particularly, the helical surface states in topological insulator accumulate an additional $\pi$ Berry phase after traveling a closed trajectory, resulting in the conductance peaks are situated at the magnetic flux of $\left(n+\frac{1}{2}\right)h/e$ (refs 26-28), which is very distinct from ordinary $h/e$ oscillations in normal metal rings[29].

Here we report the A-B oscillations in individual $Cd_3As_2$ nanowires to demonstrate the transport properties of the surface states in Dirac semimetals. A $\pi$ A-B effect that the conductance oscillations peak at odd integers of $h/2e$ with a period of $h/e$ is observed, providing transport evidence of the topological surface states of $Cd_3As_2$ nanowires. The A-B oscillations are tunable by altering the Fermi level of the nanowire via gate voltage, giving deeper insight of the 3D Dirac semimetal with Dirac fermion bulk states and Fermi arc surface states.

## Results

**Characterizations**. The $Cd_3As_2$ nanowires were synthesized *via* chemical vapor deposition (CVD) method, which are with high aspect ratio and right stoichiometry (Supplementary Fig. 1). The $Cd_3As_2$ nanowires are of single crystalline nature and grown along the [112] direction, as shown by the high-resolution transmission electron microscopy (HRTEM) image in Fig. 1a. The surface of individual nanowires is usually covered by a thin amorphous layer ~5 nm forming a core-cell structure (Supplementary Fig. 2), which may help to maintain the clean $Cd_3As_2$ surface immunized from the environment[30].

The schematic of the device and measurement configuration is shown in Fig. 1b. The temperature-dependent resistance manifests semiconducting-like behaviors (Supplementary Fig. 3), similar to other 3D Dirac semimetal materials, such as $Na_3Bi$, and is attributed to the low carrier density[17, 20]. Notable Shubnikov-de Haas (SdH) oscillations were also observed under perpendicular magnetic field (Supplementary Fig. 4). The linear extrapolation of the Landau index plot yields the intercept about 0.1, which is close to the intercept of 1/8 in 3D Dirac systems with perfect π Berry phase, giving an evidence for the Dirac fermion transport in $Cd_3As_2$ nanowires[31].

**Aharonov-Bohm oscillations.** The A-B oscillations were measured as the magnetic field was applied parallel with the longitudinal axis of the nanowire. Figure 1c shows the resistance of a nanowire with diameter ~ 115 nm (Device 1) as a function of magnetic field **B** at different temperatures. In low magnetic field regime ($|\mathbf{B}| < 0.3$T), the MR exhibits an obtuse dip due to the weak anti-localization, which is in agreement with the presence of strong spin-orbit interaction in $Cd_3As_2$ (ref. 5). At higher fields, the MR starts to oscillate periodically on a smooth negative MR background at 1.5 K. The negative MR has already been observed in Dirac semimetals and Weyl semimetals, which was believed to originate from the chiral anomaly induced charge pumping effect[17-21]. The magnitude of the negative MR is increased from -19% at 1.5 K to -34% at 20 K (Supplementary Fig. 5), which is consistent with previous report[20]. It should be noted that the (quasi-)ballistic transport along the perimeter of the nanowire is necessary to observe the A-B oscillations[28]. Therefore, the A-B oscillations are much more pronounced in the fine nanowire (~115 nm in diameter) in this work than the thick nanowire in ref. 20. As the temperature rises up to 20 K, the oscillations are nearly smeared out, indicative of its quantum nature. To focus on the quantum oscillations, the negative MR background is removed and the oscillating term $\Delta G(\mathbf{B})$ is shown in Fig. 1d. More than 10 oscillation periods (from 0 to $\pm 5$ T) can be clearly distinguished that are distributed symmetrically around the zero-field. The Fast-Fourier Transform (FFT) on $\Delta G(\mathbf{B})$ shown in Fig. 1e indicates the main peak at about $2.55\ \text{T}^{-1}$, corresponding to an individual 0.38 T period, giving a remarkable signature of flux-related quantum oscillations. If the quantum interference occurs on the surface, the *h*/*e* A-B oscillations yield a cross-section area of the nanowire about 0.11 μm². The cross-section area of the Device 1 was exactly estimated to be $0.11 \pm 0.005$μm² by direct measurements on its cut (Supplementary Fig. 6), which identifies the observed quantum oscillations are from the A-B effect.

The amplitude of oscillating term $\Delta G \sim 0.005 e^2/h$ is much smaller than the quantum conductance $e^2/h$. Previously, the observed amplitudes of A-B oscillations in topological insulator nanoribbons are also smaller than $e^2/h$ (refs 26, 28). The electrons circle along the nanowire perimeter in a quasi-ballistic manner, but drift along the longitudinal direction of the nanowire in a diffusive manner (mean free path < *L*). As probing the longitudinal conductance, the A-B oscillation amplitude of conductance may be reduced due to the diffusive transport in longitudinal direction. The amplitude of the A-B oscillation increases with decreasing temperature, as shown in Fig. 1f. It is found that the FFT peak amplitude obeys to $T^{-1}$ decay law in the temperature range of 5 to 20 K, then deviates the $T^{-1}$ dependence below ~5 K and increases more slowly as further decreasing the temperature. The $T^{-1}$ temperature dependence of oscillation amplitude indicates the nature of ballistic transport in circumferential direction[28]. The deviation from $T^{-1}$ decay below 5 K is due to the less influence of thermal broadening on the electron wave packets[32]. The critical temperature for the deviation of $T^{-1}$ dependence can be estimated by considering the critical thermal energy $E_c = \frac{h}{\pi e^2 R} \Delta E$, where $R$ is the nanowire resistance, and $\Delta E$ is the subband spacing due to geometry confinement[32]. According to nanowire boundary conditions, the subband spacing has a form of $\Delta E = \frac{h v_F}{W} = h v_F \sqrt{\frac{1}{4\pi S}}$, where $W$ is the perimeter of nanowire, $S$ is cross-sectional area and $v_F$ is Fermi velocity[26]. Hence, the critical temperature is calculated to be ~7.2 K, which is close to the experimental observation of ~ 5 K.

The elastic scattering length $L_0$ ~ 10 nm of the longitudinal transport along the nanowire is estimated by the amplitude of oscillating term $\Delta G(\mathbf{B}) \sim 0.005 e^2/h$ *via* the $Landauer-Büttiker$ formula of $G = \frac{e^2/h}{1+L/L_0}$ (ref. 33), where $L \sim 2\mu m$ is the distance between the two voltage probes. Although the electrons circle along the nanowire perimeter in a quasi-ballistic manner, the carrier drift along the longitudinal direction of the nanowire is in a diffusive manner (mean free path < $L$). Therefore, the obtained $L_0$ can be underestimated due to the reduced amplitude $\Delta G$. Nevertheless, the ballistic or quasi-ballistic transport along the nanowire perimeter must be satisfied to observe the A-B oscillations[26-28,30,34]. Therefore, it is naturally inferred that the nontrivial surface states in the $Cd_3As_2$ nanowire should have high carrier mobility with suppressed scatterings by topological protection. Our results are also consistent with the observations in topological insulator nanowires that are not ballistic along the nanowire length but the ballistic along the nanowire perimeter[34].

The oscillations in $\Delta G$ as a function of magnetic flux $\mathbf{\Phi}$ in unit of *h/e* are presented in Fig. 2a. In the two initial periods, the conductance peaks are at the integer quantum flux ($\mathbf{\Phi} = h/e$, $2h/e$). Then, a phase-shift takes place in the range from 1 to 2 T, wherein the conductance peaks are at random flux $\mathbf{\Phi} = 3.1h/e$ and $4.3h/e$. After the phase-shift region, the conductance peaks are stable at the half-integer quantum flux ($\mathbf{\Phi} = (5+1/2)h/e$, $(6+1/2)h/e$, …). The conductance oscillation peaks at half-integer of $\mathbf{\Phi}_0$ with a period of $\mathbf{\Phi}_0$ are very similar with that observed in topological insulators, which was ascribed to the spin-related $\pi$ Berry phase of the topological surface states[24,25]. At $\mathbf{\Phi}/\mathbf{\Phi}_0 = 0$, the inter-spacing between subbands due to the finite boundary condition is estimated by $\Delta E = hv_F/W$ (refs 26-28). At $\mathbf{\Phi}/\mathbf{\Phi}_0 = n$, the surface state is fully gapped at $E = 0$. While at $\mathbf{\Phi}/\mathbf{\Phi}_0 = n+1/2$, a gapless surface band appears due to the additional $\pi$ Berry phase[26]. Intuitively, to observe this anomalous A-B oscillation, it requires that the Fermi-level $E_F$ is close to $E=0$ or the quantum confinement induced gap is large enough. Due to the high crystal quality with fewer defects, the $Cd_3As_2$ nanowires are of low carrier density with the order of ~$10^{17}$ cm$^{-3}$, which has been measured by field-effect gating method and found to be the crucial factor for the observation of the negative MR effect in our previous work[20]. The low carrier density also suppresses the bulk conductance with the resistance $R > 25.8$ kΩ at low temperatures, promoting the detection of the surface states of the nanowire with large surface-to-volume ratio.

To understand the phase-shift of the A-B effect in $Cd_3As_2$ nanowires, we would like to discuss the origin of the additional $\pi$ Berry phase by considering the evolution of Dirac point under magnetic fields. In $Cd_3As_2$, the Dirac points are distributed equally away from the Γ point along the [001] direction at $\pm k_D$ (ref. 5). As schematically shown in Fig. 2b, without magnetic field, the Fermi arc surface states connect the two bulk Dirac nodes and each Dirac node contains two degenerate Weyl nodes. Applying an external magnetic field can break the time reversal symmetry and split the two Dirac points into two pairs of Weyl nodes along to the magnetic field direction[35]. As schematically shown in Fig. 2c, under magnetic field, each surface Fermi arc connects the Weyl nodes with right handed chirality and left handed chirality, respectively. The chirality-nontrivial surface states are non-degenerate. The lifting of electronic state degeneracy results in an additional Berry phase $\pi$ for cyclotron electrons after cycling around one of the Weyl nodes in the low energy band. Moreover, the analysis of SdH oscillations also suggests the presence of non-trivial $\pi$ Berry phase (Supplementary Fig. 4)[16]. It is very interesting to compare our results with the A-B oscillations in topological insulator nanoribbons. The surface subbands have the 1D band dispersion as $E(n,k,\Phi) = \pm h v_F [\frac{k^2}{4\pi^2} + (n + 0.5 - \frac{\Phi}{\Phi_0})^2 / L^2]^{1/2}$, where the A-B oscillation in conductance is anticipated to

7 / 31

peak at odd integers of $\Phi_0/2$ (refs 26-28). Similarly, the A-B oscillation with peaks at odd integers of $\Phi_0/2$ was also found in our $Cd_3As_2$ nanowires under a magnetic field larger than a critical value. This phase-shift is believed to be in close relationship to the extra $\pi$ Berry phase released by the magnetic field induced lifting of degeneracy.

**Angular dependence of A-B oscillations.** To verify the influence of magnetic flux on the A-B effect, the nanowire direction was tilted, forming an angle $\theta$ with the direction of the magnetic field (Fig. 3a). As shown in Fig. 3b, the magneto-transport results illustrate a negative to positive MR transition from $\theta = 0^o$ (parallel case) to $\theta = 40^o$. The oscillations in $\Delta G$ obtained at different tilted angles are presented in Fig. 3c. The weak anti-localization induced conductance peak at **B**=0 is not sensitive to the tilt angle, indicating its bulk origin. At low magnetic field, the periodic oscillations at high tilt angles are still discernible. By scaling the magnetic field with $B\cos\theta$, we replot the oscillations at different tilt angles in Fig. 3d. It is found that the oscillation peaks are line up in Fig. 3d, suggesting the dominant magnetic flux along the nanowire longitudinal direction. Meanwhile, the oscillation peaks split at high tilt angles, which may be due to the deflection of the electron trajectory under the Lorentz force from the perpendicular component $B_\perp$ of the magnetic field. It is also found that the periodic oscillations vanish gradually as increasing $\theta$ under high magnetic fields. The vanishing of oscillations under high magnetic field may be due to the $B_\perp$ induced breaking of the time reversal symmetry of the surface states and a bandgap opening, which is consistent with that observed in topological insulators[26].

**Gate-voltage dependence of A-B oscillations.** The periodic conductance oscillation is caused by the rearrangement of the surface subbands in energy as varying the magnetic flux[26, 28]. Therefore, the occupation of subbands can affect the pattern of oscillation in aspect of magnitude and phase. By employing a back-gate with 285 nm SiO$_2$, we are able to modulate the subband occupation by applying gate voltage ($V_g$), and explore how the Fermi level affects the A-B oscillations. To achieve this goal, we performed systematic gate-modulation experiments on a very thin Cd$_3$As$_2$ nanowire with diameter ~ 65 nm (Device 2), in which a large subband spacing $\Delta E = \frac{hv_F}{W} \sim 21 \text{meV}$ can be obtained. As shown in Fig. 4a, the transfer curve indicates the ambipolar field effect of the device. The resistance saturates to a certain value at high gate-voltages (> 5 V), indicating that the screening effect may hinder further gate-modulating as the Fermi-level is far away from the Dirac point. Nevertheless, as shown in Fig. 4a, the resistance can be largely tuned at small gate voltages $V_g$, demonstrating the effective gate-modulation as long as the screening effect is not strong enough.

We have measured the magnetotransport properties of the nanowire in the gate voltage range without obvious screening effect. As shown in Fig. 4b, the A-B oscillations in $\Delta G$ were plotted as a function of magnetic flux at different $V_g$ of -7.5 V, -3 V, 0 V, 3 V and 5 V. It is worth noting that recent studies on Cd$_3$As$_2$ thin films (~50 nm in thickness) demonstrate that a considerable band gap >24.9 meV may be opened at the Dirac point as a result of quantum confinement effect[36]. Because this nanowire is with diameter of ~ 65 nm, it is very likely that a band gap is also opened at the Dirac point in the nanowire. Therefore, at $V_g$ = 0 V (near the Dirac point as seen in Fig. 4a), the Fermi level is in the band gap. As the Fermi level is inside the bandgap, the A-B oscillation thus has a normal origin, which may be responsible for the observation of the maxima of $\Delta G$ at even integers of $\Phi_0/2$ at $V_g$ = 0 V, as shown in Fig. 4b. As tuning the Fermi level into the valence band, there is a phase shift of the A-B oscillations at $V_g$ = -3 V. As the Fermi level further enters into the valence band with linear dispersion, the π-A-B effect is observed at $V_g$ = -7.5 V, as shown in Fig. 4b, where the maxima of Δ$G$ at odd integers of $\Phi_0/2$.

As tuning the Fermi level into the conduction band, the A-B oscillations are still clearly observed but with more complicated features at $V_g$ = 3 V and 5 V, as shown in Fig. 4b. Previous studies indicate that the electron has much higher mobility than that of hole in $Cd_3As_2$ due to the different Fermi velocity and scattering rate with defects [9, 14, 37]. The magnetic field can force the bulk electrons to the surface under the classic limit of $R < L_B = \sqrt{\hbar/eB}$, *i.e.*, the nanowire radius $R$ is smaller than the magnetic length $L_B$ (refs 38, 39). The electrons with high carrier mobility deflected from the bulk channel contribute to the surface conductance significantly, resulting in the irregular A-B oscillations. Furthermore, we have performed more systematic gate-dependent A-B oscillations on another sample with similar size (radius ~ 37 nm, Device 3), as shown in Supplementary Fig. 7. The oscillation patterns are complicated at relatively weak magnetic fields, implying the co-contribution to the quantum oscillations from both the bulk and surface electrons. Fortunately, further increasing $B$ may give rise to less scattering on the surface from bulk electrons as $L_B < R$, and there is a clear π-A-B effect under high magnetic field (Supplementary Fig. 7b).

In Dirac semimetals, the two bulk Dirac points are separated by $2k_D$ in momentum space. The relative magnitude of $k_D$ and the Fermi wave vector $k_F$ is thus crucial to describe the electronic structure in Dirac semimetals. Further increasing $k_F$ larger than $k_D$, two Fermi pockets nested at $\pm k_D$ merge into one entire Fermi surface centered at Γ, called the Lifshitz transition[14]. As the system is back to degenerate system, the π Berry phase should disappear. To verify this scenario, we measured a nanowire (Device 4) with back gate voltage up to 60 V. At $V_g$ = 60 V, the oscillation peaks at relatively random positions under low magnetic fields (Supplementary Fig. 8), which may be due to the large magnetic length of the electron orbit and high carrier density. In this situation, the bulk carrier can be scattered to the surface, which makes the A-B oscillations complicated. As increasing magnetic field, the magnetic length of the bulk electron cyclotron motion decreases and the oscillation starts to peak exactly at the integer quantum flux with $\Phi$ = 5, 6, 7, 8 $h/e$ (Supplementary Fig. 8). Since the Fermi level is far from the Dirac points, the Fermi surface may enter above the Lifshitz transition, thus leading to the normal A-B oscillations.

**Altshuler–Aronov–Spivak effect.** For the thick nanowire device, the electron transport along the nanowire perimeter may be in the diffusive regime. Figure 5a shows a thick nanowire with diameter about 200 nm (Device 5). Even in a relatively narrow field range from -1 to 1 T, prominent oscillations superposed on negative MR background can be easily discerned, as shown in Fig. 5b. According to the nanowire cross-sectional area (Supplementary Fig. 6), it is found that there are oscillations with $h/2e$ flux period, as noted by arrows in Fig. 5c. These $h/2e$ oscillations may be attributed to the Altshuler–Aronov–Spivak (AAS) effect in diffusive regime. Notably, the $h/2e$ oscillation peaks at $\left(n + \frac{1}{2}\right) h/2e$ ($n$=2,3,4), indicating a phase shift for the $h/2e$ oscillations that needs further investigations.

**Discussion**

In summary, we have studied the A-B effect in individual Dirac semimetal $Cd_3As_2$ nanowires as the magnetic flux penetrates through the cross-section of the nanowire. As increasing magnetic field, a phase shift in A-B oscillations was observed, which is ascribed to the splitting of the Dirac node into Weyl nodes due to the magnetic field induced time-reversal symmetry breaking. With this lifting of the Dirac node degeneracy, the conductance oscillation peaks at odd integers of $h/2e$ with period of $h/e$, which provides transport evidence of the topological surface states of $Cd_3As_2$ nanowires. As tuning the Fermi level by gate voltage, a transition between the 0 A-B effect and the $\pi$ A-B effect was observed, which is self-consistent with the presence of linear dispersion relation in 1D-like quantum confined system. In thick nanowires, the A-B oscillations were largely reduced and the AAS oscillations presented due to the transition from quasi-ballistic to diffusive transport. The A-B oscillations observed in the Dirac semimetal nanowires show abundant features, offering a route to explore the exotic surface states.

## Methods

**Growth of the nanowires.** The $Cd_3As_2$ nanowires were synthesized by CVD method in a horizontal furnace. For safety, the whole system was placed in a ventilation closet and the pumper was equipped with a cold trap to collect the dust in the exhaust. The $Cd_3As_2$ powders (purity>99.99%, Alfa Aesar) placed at the center of the tube furnace were heated from room temperature to 650 ℃ in 20 min after pumping and flushing the tube several times to get rid of oxygen. The Si substrates covered by a 5 nm thick gold film were placed at the downstream away from the source ~ 15 cm to collect the products. The system was maintained at 650 ℃ for 10 min and at about 1 atm pressure with about 20 s.c.c.m Argon flow during the growth process. Then, the system cooled down to room temperature naturally.

**Device fabrication.** The $Cd_3As_2$ nanowire devices were fabricated on a Si substrate with 285 nm $SiO_2$ layer. To make the nanowire direction can be parallel with the direction of the magnetic field, micro-strips were first prefabricated on a SiO$_2$/Si square substrate as marks, which are parallel to one side of the square substrate. Then, the nanowire was transferred from the as-grown substrate to the SiO$_2$/Si square substrate, and the nanowire direction was aligned parallel to the prefabricated marks by a micromanipulator. The individual nanowire devices were fabricated using standard electron beam lithography techniques. To form Ohmic contacts between Cd$_3$As$_2$ nanowires and electrodes, the amorphous layer of the nanowire was removed *in-situ* by Ar$^+$ etching treatment in the metal deposition chamber. Then, ~ 150 nm thick Au electrodes were deposited. The linear current-voltage curve (Supplementary Fig. 9) indicates the Ohmic contacts between the nanowire and the electrodes.

**Transport measurements.** The SiO$_2$/Si square substrate with the Cd$_3$As$_2$ nanowire devices was mounted onto the sample holder in an Oxford commercial variable temperature insert to make the nanowire parallel to the direction of the magnetic field. The sample holder can be rotated continuously by an electric motor to change the angle between **E** and **B**. The electrical signals were measured using a four-terminal method by Stanford SR830 lock-in amplifiers at frequency of 17.7 Hz.

## Acknowledgements

We are grateful to Prof. Xincheng Xie, Prof. Ji Feng, Prof. Kai Chang, and Dr. Haiwen Liu for inspired discussions. This work was supported by MOST (Nos. 2013CB934600, 2013CB932602) and NSFC (Nos. 11274014, 11234001, 11327902).

## Author contributions

Z.-M.L. conceived and designed the experiments. D.-P.Y. gave scientific advice. L.-X.W. and C.-Z.L. performed the measurements. Z.-M.L. and L.-X.W. wrote the manuscript.

**Competing financial interests:** The authors declare no competing financial interests.

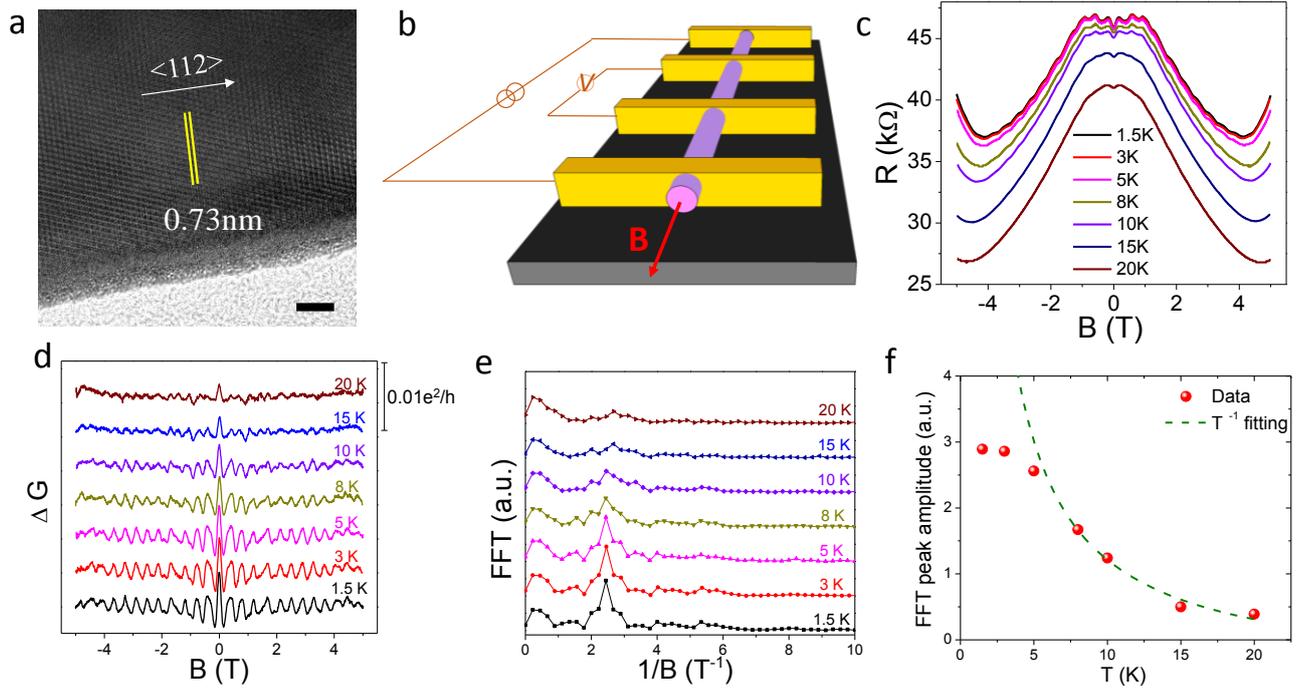

**Figure 1 | Magnetotransport of a nanowire with diameter ~ 115 nm (Device 1).** (**a**) HRTEM image of a typical nanowire indicates <112> growth direction with interplanar space ~ 0.73 nm. Scale bar: 5 nm. (**b**) Schematic diagram of the four-terminal device with applied magnetic field aligned with the length. (**c**) Resistance as a function of magnetic field at different temperatures from 1.5 K to 20 K. (**d**) The oscillations in conductance as a function of magnetic field after subtracting background. (**e**) FFT spectrums of the conductance oscillations. (**f**) Plot of the temperature dependence of the FFT peak amplitude. The dashed line indicates the fitting by the $T^{-1}$ dependence.

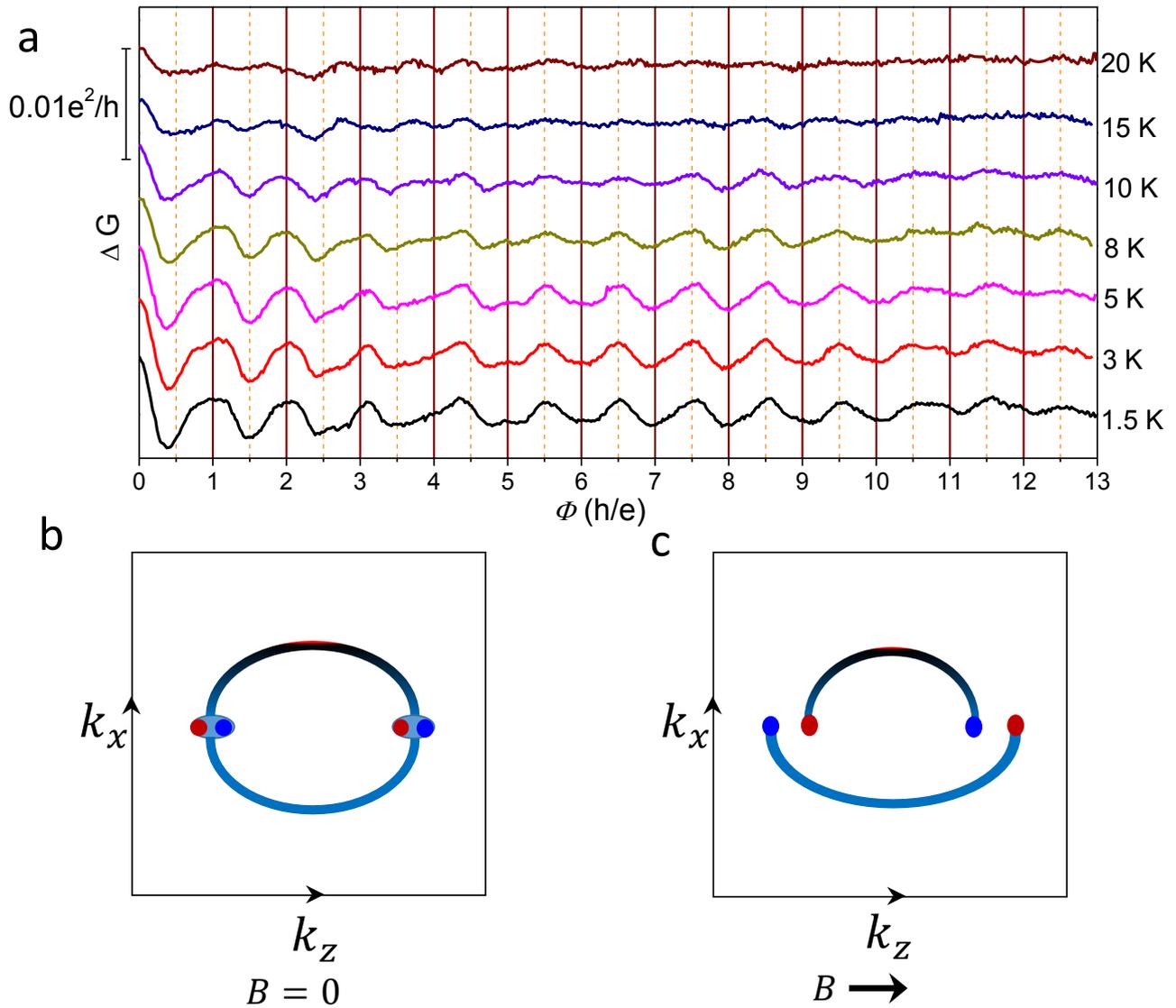

**Figure 2 | Aharonov-Bohm oscillations.** (**a**) Oscillating conductance as a function of magnetic flux in unit of *h/e* at variable temperatures. (**b,c**) Schematic diagram of the Weyl nodes projected onto the 2D $k_x - k_z$ surface plane (**b**) without and (**c**) with an external magnetic field. The right handed chirality and left handed chirality are denoted by the red ball and blue ball, respectively. The Fermi arcs connect the Weyl nodes with opposite chirality. Without magnetic field, each Dirac node consists of two degenerate Weyl nodes. A magnetic field can separate the two degenerate Weyl nodes in the momentum space.

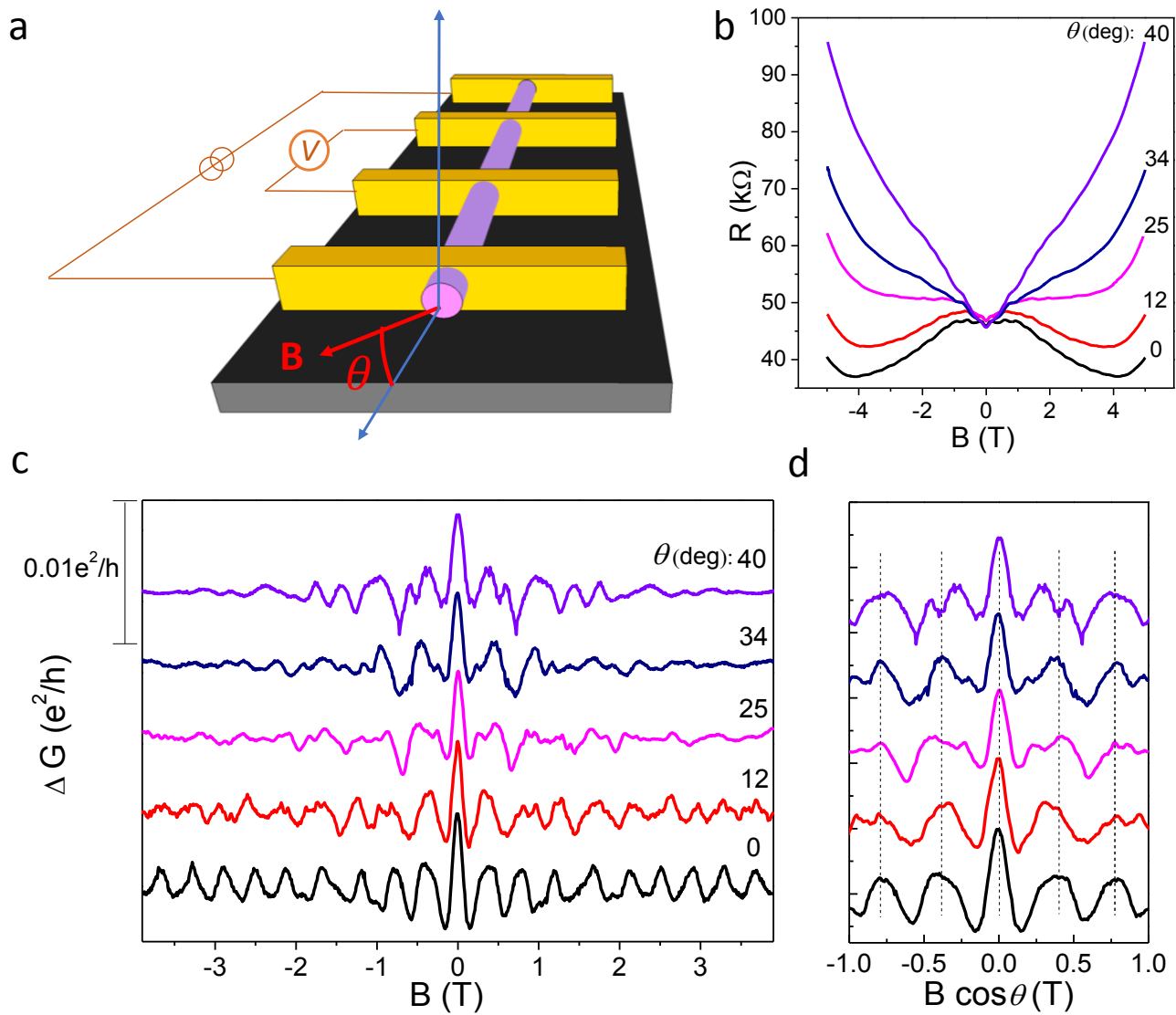

**Figure 3 | Aharonov-Bohm oscillations at tilted magnetic field.** (**a**) Schematic of the measurement configuration. The $\theta$ is the angle between the orientation of the magnetic field and the longitudinal direction of the nanowire. (**b**) Magnetoresistance at different $\theta$ angles and (**c**) the corresponding conductance oscillations. (**d**) The conductance oscillations as a function of $B\cos\theta$.

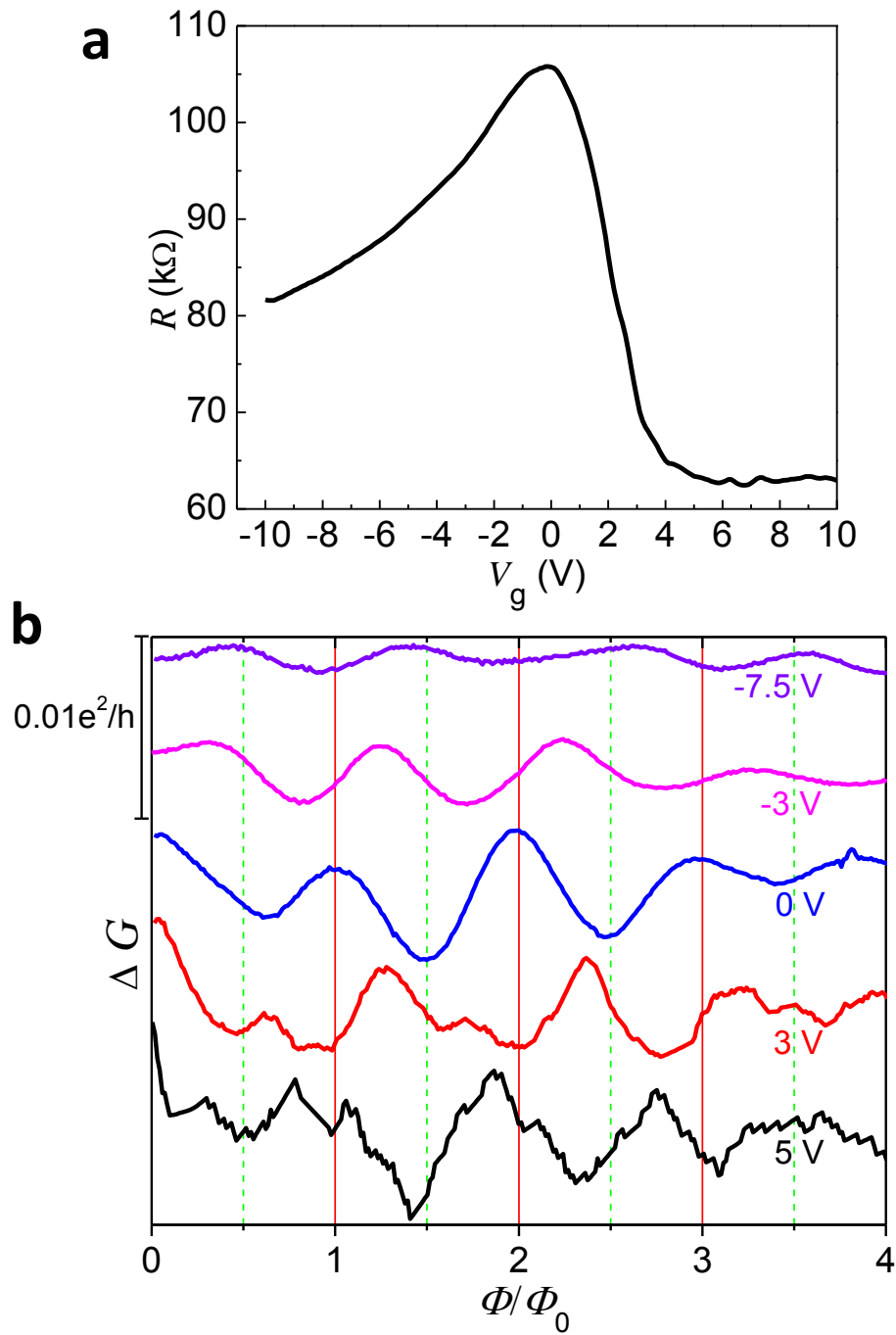

**Figure 4 | Gate tuning of Aharonov-Bohm oscillations.** (**a**) Gate voltage dependent resistance of a nanowire with diameter ~ 65 nm (Device 2) at 1.5 K. (**b**) A-B oscillations at gate voltages as denoted and at 1.5 K. The positive and negative gate voltages pull the Fermi level into the conduction band and valence band, respectively.

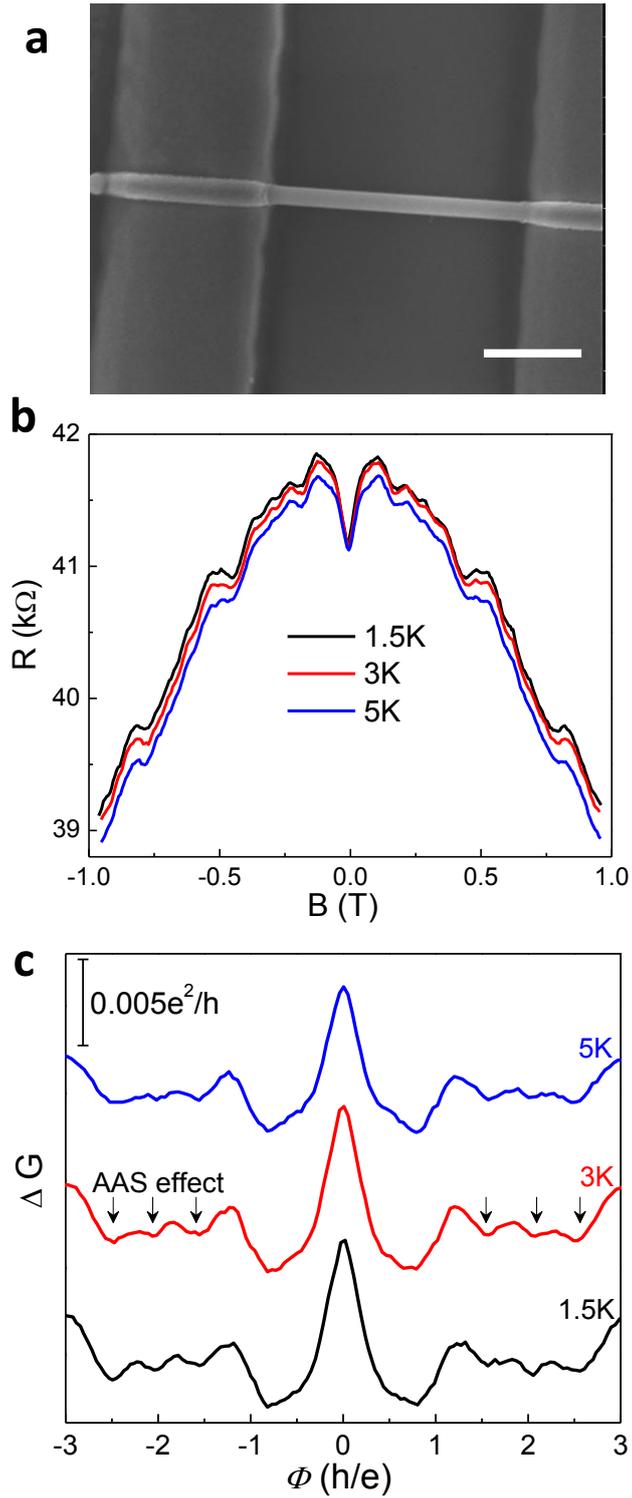

**Figure 5 | Altshuler–Aronov–Spivak oscillations.** (**a**) SEM image of a thick nanowire with diameter about 200 nm (Device 5). Scale bar: 1 μm. (**b**) Magnetoresistance at different temperatures. (**c**) The conductance oscillations as a function with magnetic flux, showing the AAS effect.

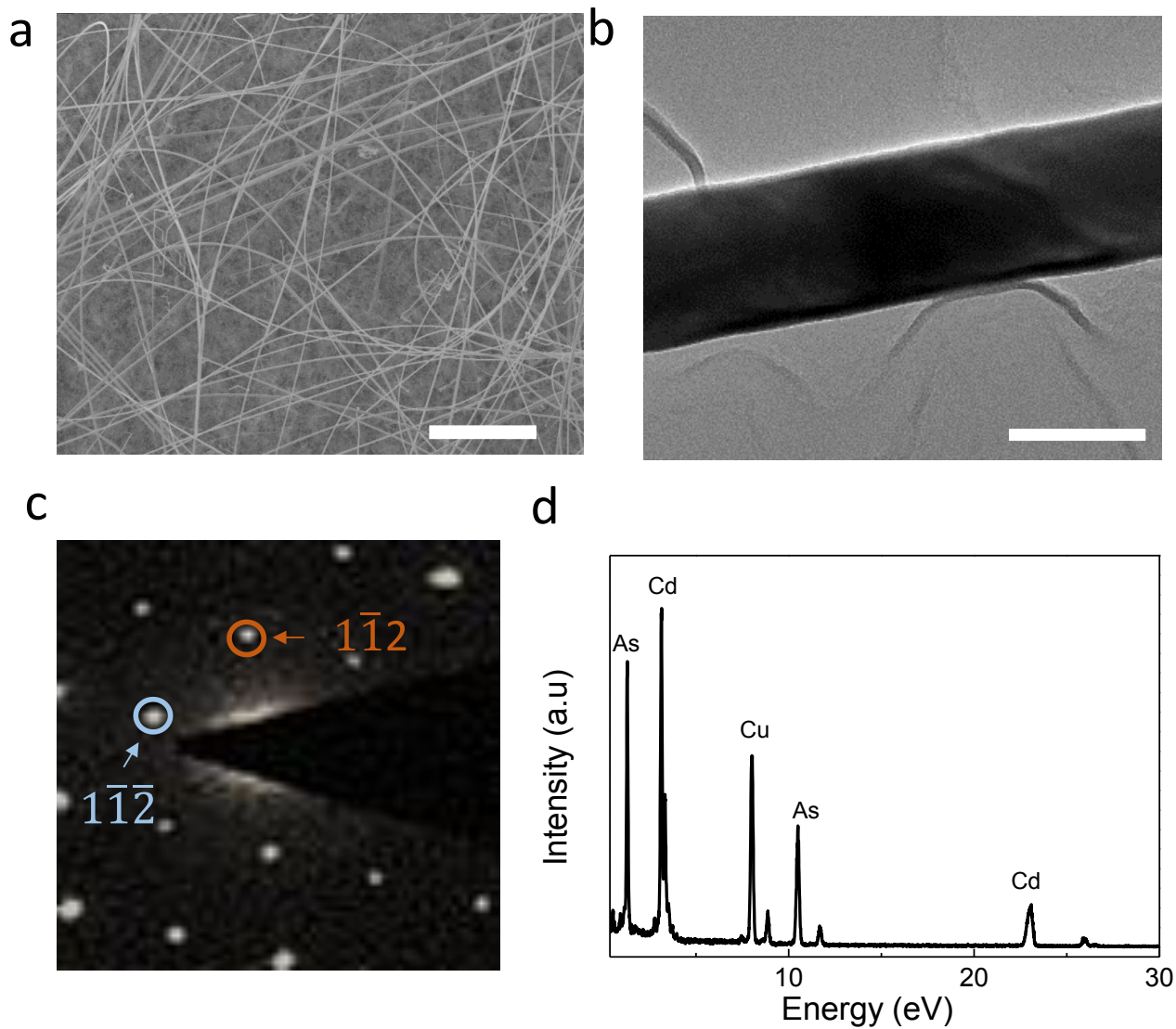

**Supplementary Figure 1 | Characterization of Cd₃As₂ nanowires.** (**a**) SEM image (Scale bar: 20 µm) and (**b**) TEM image (Scale bar: 100 nm) of Cd$_3$As$_2$ nanowires. (**c**) Typical electron diffraction pattern and (**d**) the energy-dispersive X-ray spectroscopy (EDS) results of Cd$_3$As$_2$ nanowires. The as-grown Cd$_3$As$_2$ nanowires manifest great flexibility. The nanowire diameter is in the range from tens of nanometers to several hundreds of nanometers. The selected electron diffraction pattern in (**c**) indicates the growth direction is along the [112] crystalline axis. The EDS results in (**d**) show the right stoichiometry of the Cd$_3$As$_2$ compound.

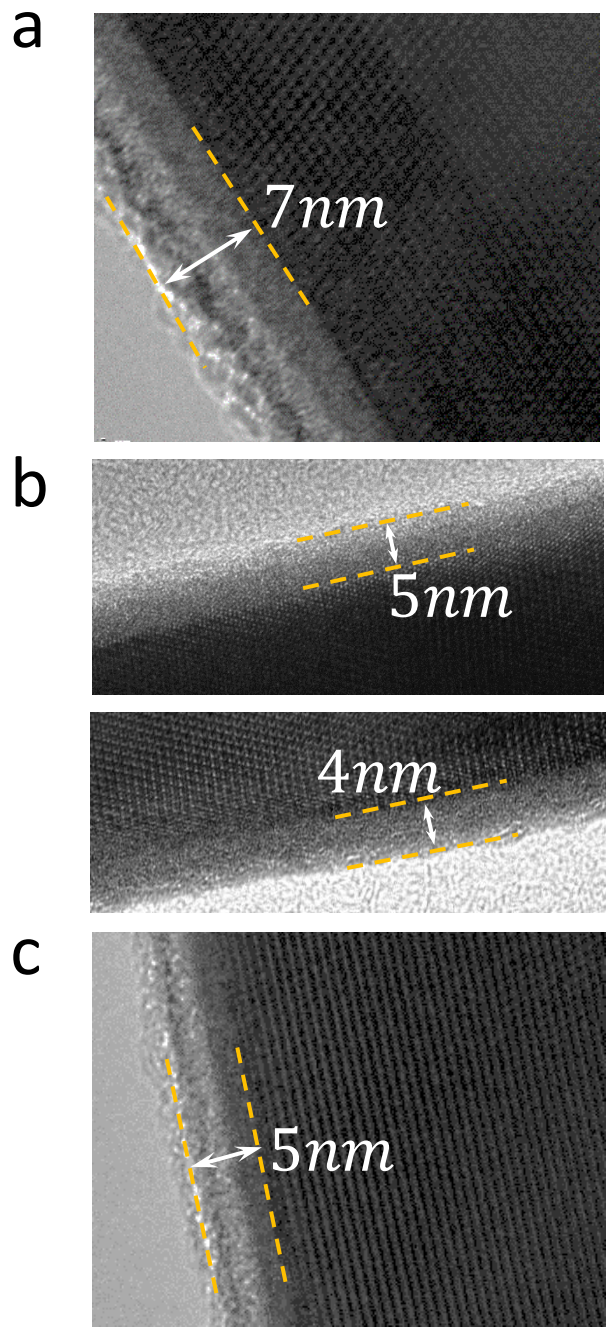

**Supplementary Figure 2 | Amorphous layer on the nanowire surface.** High resolution TEM images of several typical $Cd_3As_2$ nanowires. Amorphous layer on $Cd_3As_2$ nanowires is generally observed in the TEM images, which may be due to the natural oxidation of the surface after exposed in the air.

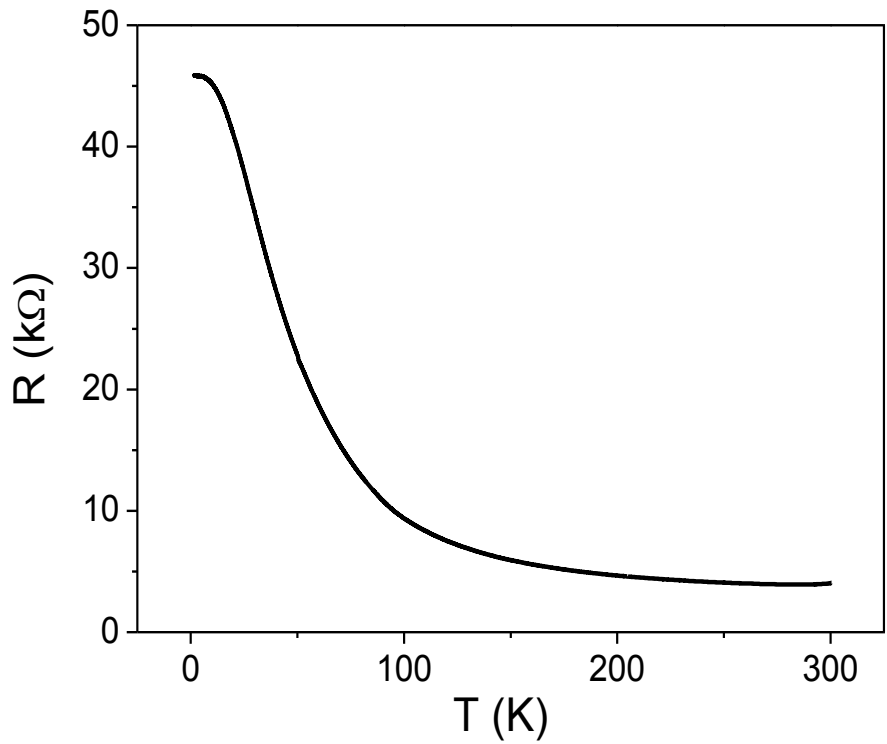

**Supplementary Figure 3 | Resistance as a function of temperature.** The semiconducting-like *R-T* behavior indicates that the Fermi level is close to the Dirac point due the low carrier density of the nanowire. The thermal energy can activate the carriers to enhance the conductance, resulting in the resistance decreases with increasing temperature.

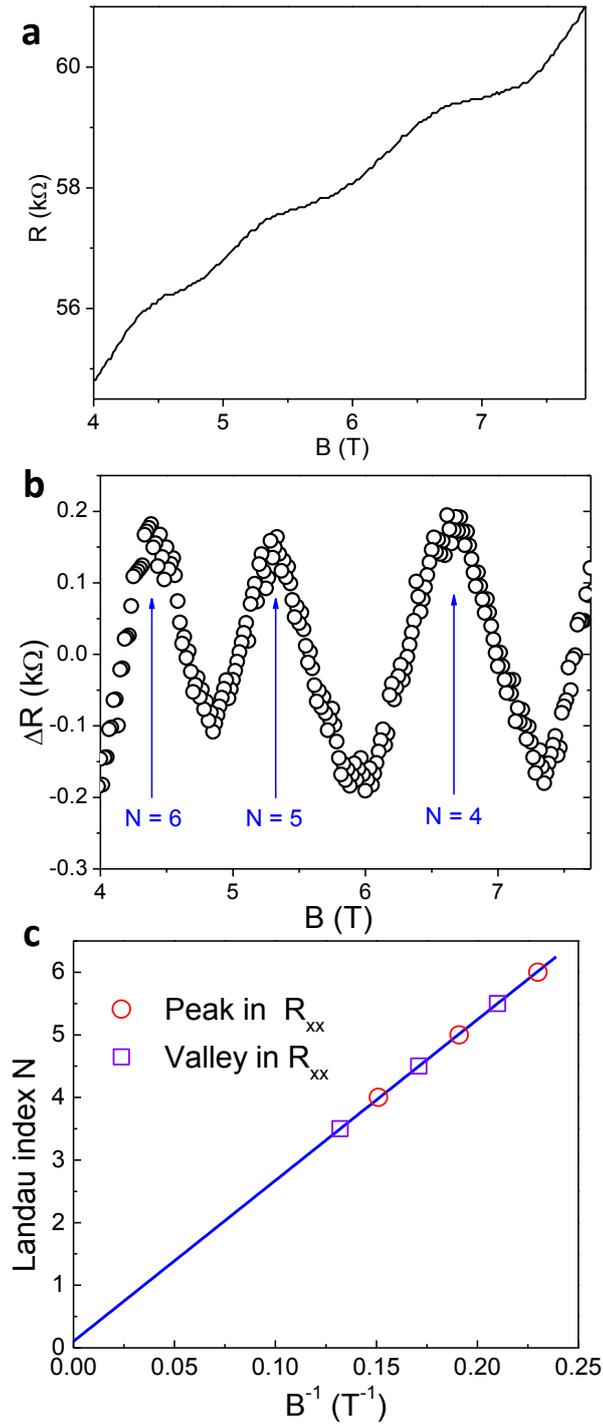

**Supplementary Figure 4 | Shubnikov-de Haas oscillations.** (**a**) The SdH oscillations in the plot of resistance *R* *vs.* magnetic field *B*. (**b**) Extracted oscillation part $\Delta R$ as a function of magnetic field *B*. (**c**) Landau index plot by assigning integers to the maxima in $\Delta R$, yielding a intercept of 0.1 by the linear extrapolation.

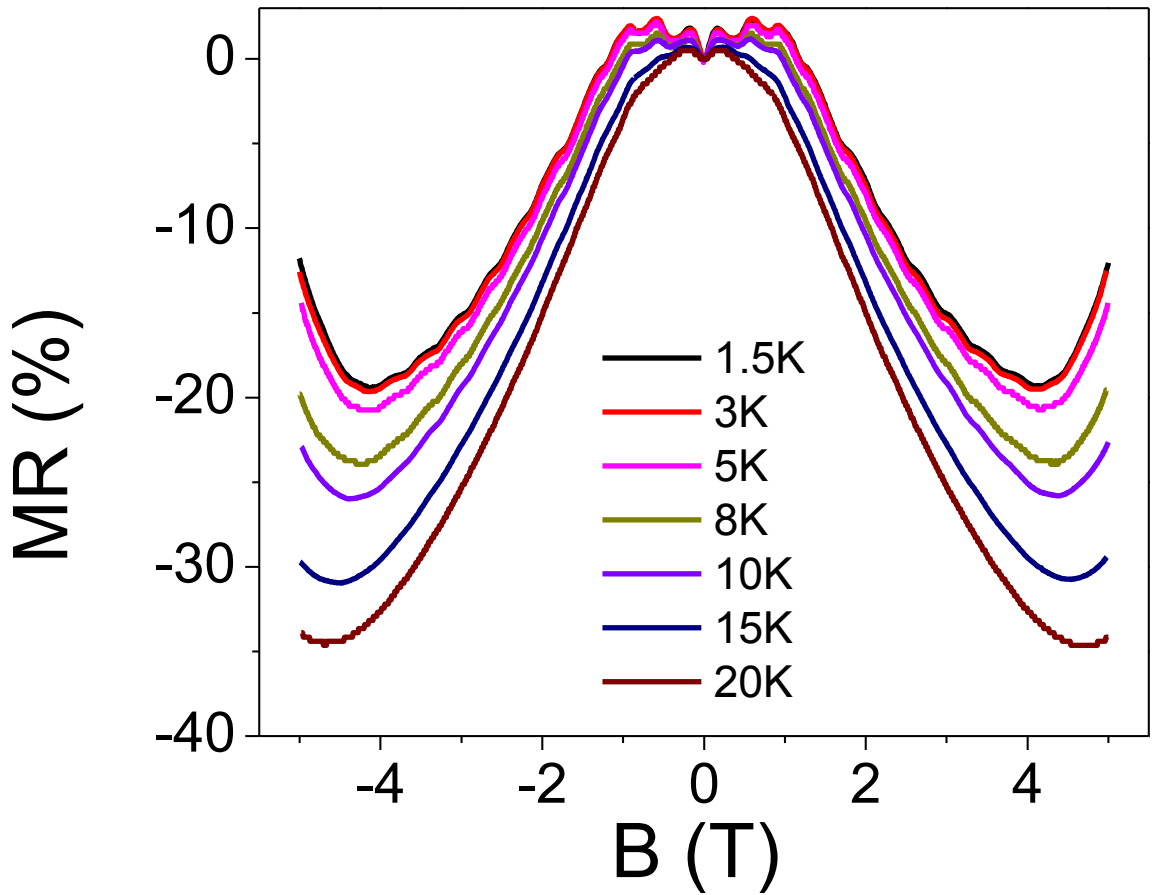

**Supplementary Figure 5 | Negative magnetoresistance.** Magnetoresistance, defined as MR=[$R(B)/R(0)$-1]×100%, of $Cd_3As_2$ nanowires at variable temperatures in $B//E$ configuration shows negative values, giving signature of the chiral anomaly effect.

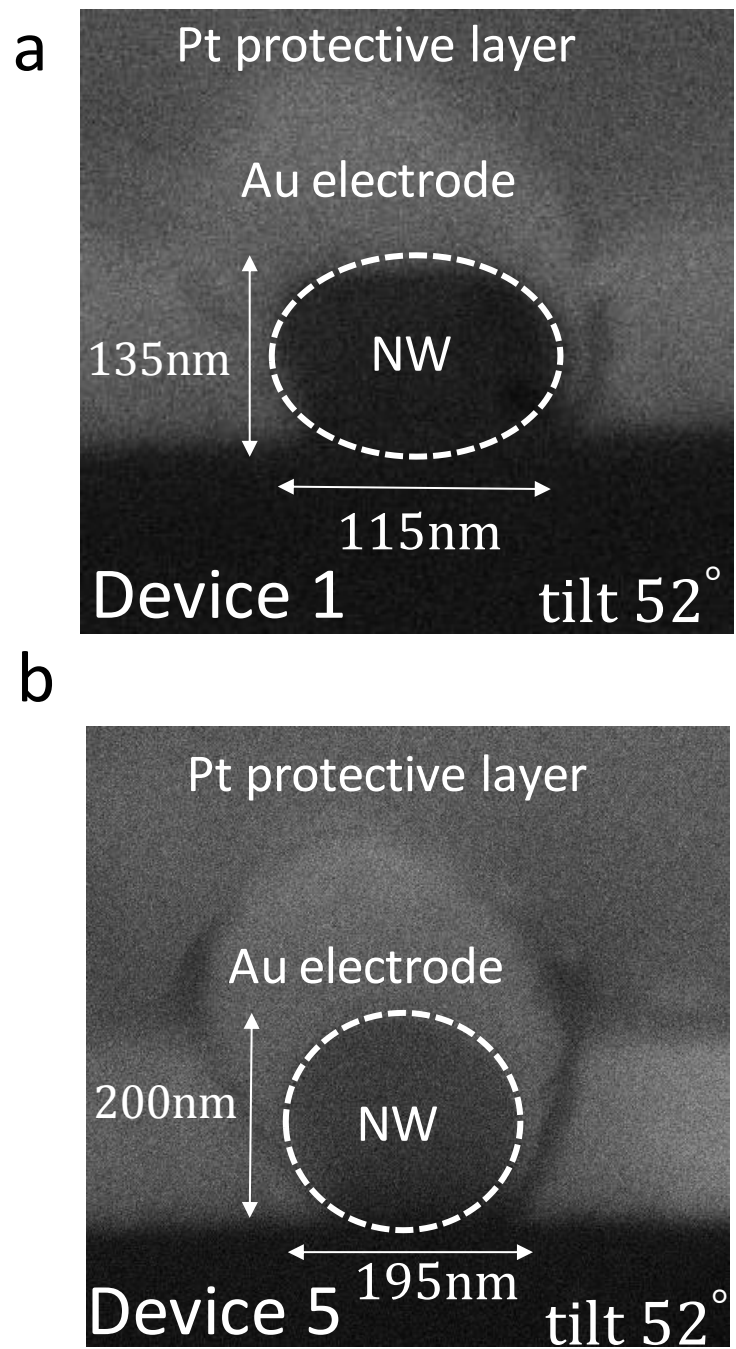

**Supplementary Figure 6 | Cross-section of Devices 1 and 5.** SEM images at tilt angle 52° of nanowire cross-section of (**a**) Device 1 and (**b**) Device 5. To acquire the cross-sectional area of the nanowire devices, a thin Pt film was first deposited on the nanowire as a protective layer by electron beam induced deposition method in a dual beam system (Dual-Beam 235-FIB system, FEI Company). Then, the nanowires were cut by focused ion beam.

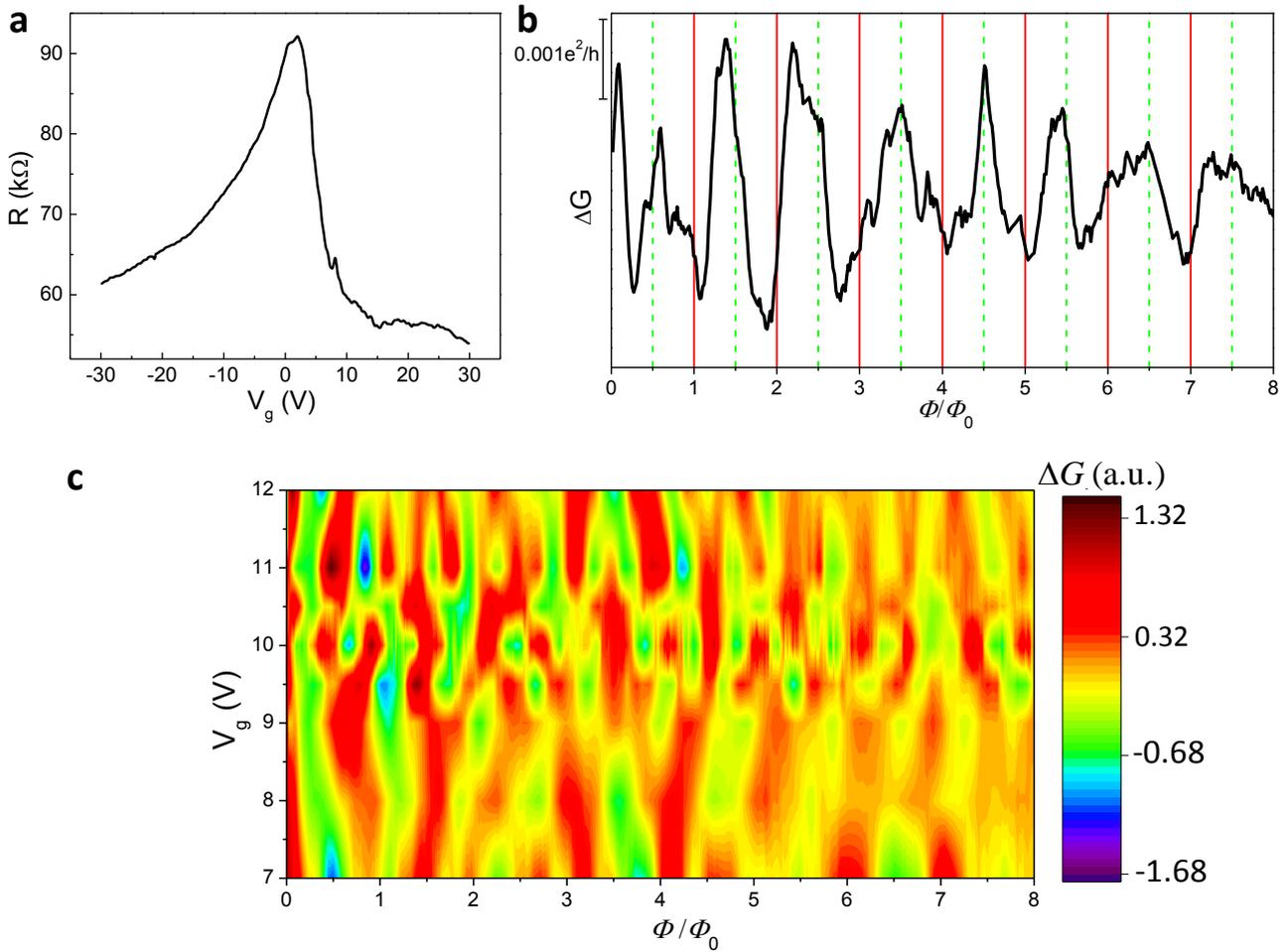

**Supplementary Figure 7 | Gate modulated transport properties of a nanowire with radius of ~37 nm (Device 3).** (**a**) Transfer curve at 1.5 K. (**b**) Oscillation part $\Delta G$ as a function of magnetic flux in unit of $\Phi_0 = h/e$ at $V_g$ = 10.5 V. (**c**) Mapping of gate-dependent conductance oscillations from $V_g$ = 7 to 12 V with a step of 0.5 V. For comparation, the $\Delta G$ was normalized.

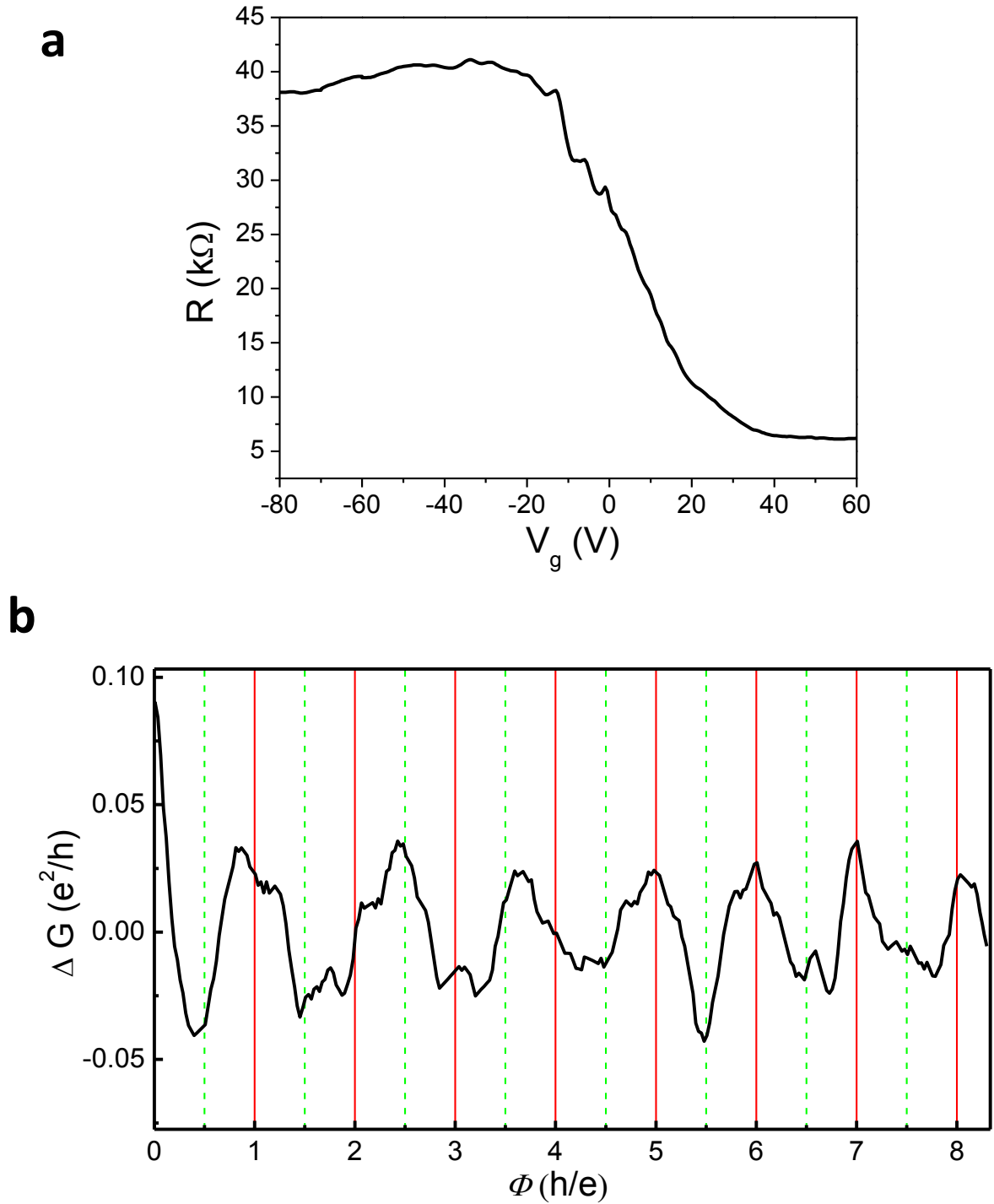

**Supplementary Figure 8 | Gate modulated transport properties of a nanowire with diameter of ~90 nm (Device 4).** (**a**) Resistance as a function of applied gate-voltage $V_g$ at 1.5 K. (**b**) A-B oscillations as a function of $\Phi$ at $V_g = 60$ V after subtracting the background.

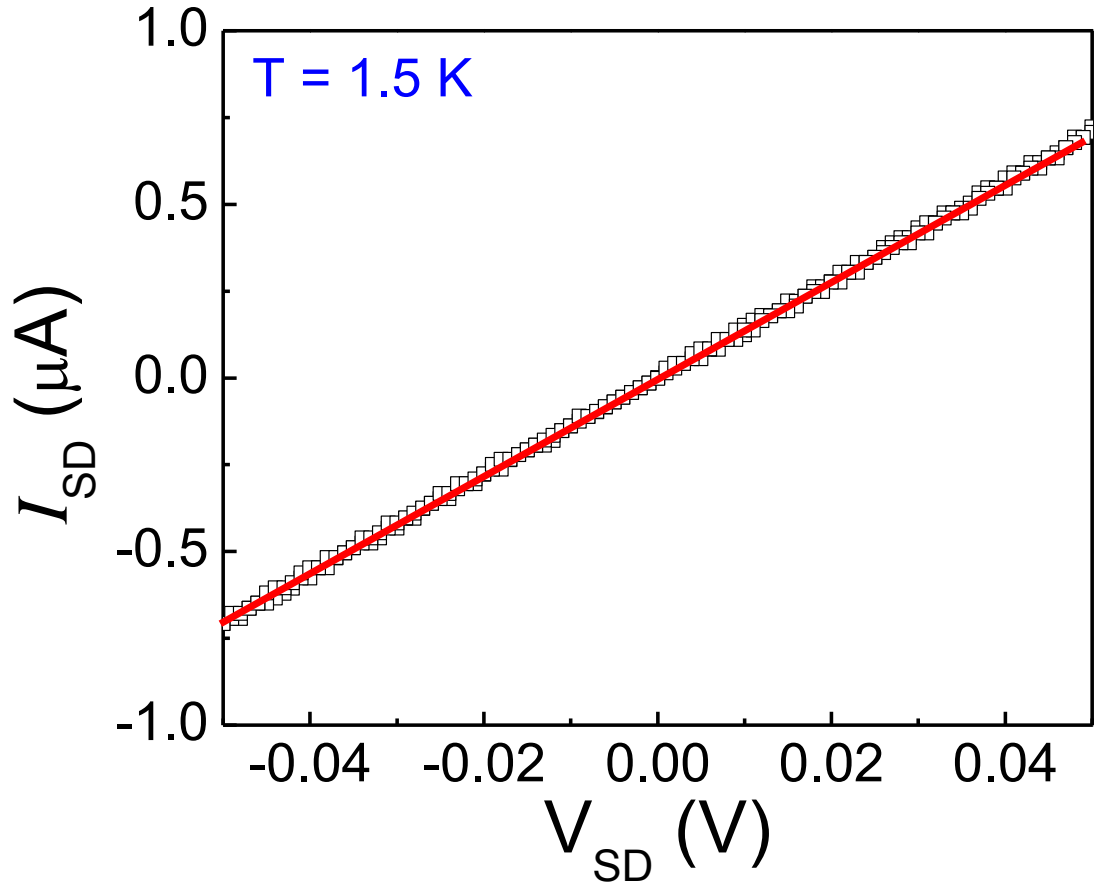

**Supplementary Figure 9 | Current-voltage curve of Device 5 at 1.5 K.** The linear source-drain current $I_{SD}$ vs. source-drain voltage $V_{SD}$ curve indicates the Ohmic contacts between Au electrodes and $Cd_3As_2$ nanowires.